
\input phyzzx
\nonstopmode
\sequentialequations
\twelvepoint
\nopubblock
\tolerance=5000
\overfullrule=0pt
\input epsf

\line{\hfill PUPT 1550, IASSNS 95/59}
\line{\hfill cond-mat/9507016}
\line{\hfill May 1995}
\titlepage
\title{Spin-Singlet to Spin Polarized Phase Transition at $\nu=2/3$:
Flux-Trading in Action}
\vskip.2cm
\author{Chetan Nayak\foot{Research supported in part by a Fannie
and John Hertz Foundation fellowship.~~~
nayak@puhep1.princeton.edu}}
\vskip .2cm
\centerline{{\it Department of Physics }}
\centerline{{\it Joseph Henry Laboratories }}
\centerline{{\it Princeton University }}
\centerline{{\it Princeton, N.J. 08544 }}

\author{Frank Wilczek\foot{Research supported in part by DOE grant
DE-FG02-90ER40542.~~~WILCZEK@IASSNS.BITNET}}
\vskip.2cm
\centerline{{\it School of Natural Sciences}}
\centerline{{\it Institute for Advanced Study}}
\centerline{{\it Olden Lane}}
\centerline{{\it Princeton, N.J. 08540}}
\endpage

\abstract{We analyze the phase transition between spin-singlet
and spin-polarized states which occurs at $\nu=2/3$.
The basic strategy is to use adiabatic flux-trading arguments
to relate this transition to the analogous transition at
$\nu=2$. The transition is found to be similar to a transition
in ferromagnets. In our analysis, we find two possible scenarios.
In one, the transition is first-order,
in agreement with experimental and numerical studies of the
$\nu=2/3$ transition. In the other, we find a
second-order transition to a partially polarized state
followed by a second-order transition to
a fully polarized state. This picture is in qualitative
agreement with experiments on the $\nu=4/3$ state,
the particle-hole conjugate of $\nu=2/3$.
We analyze the edge
modes which propagate at the boundaries between regions
of different phases and show that these do not
support gapless excitations. Finally, we consider the possibility
of a finite-temperature compressible state with a Fermi surface
which would explain the non-zero $\rho_{xx}$
seen in experiments.}

\endpage

\REF\halperin{B.I. Halperin, Helv. Phys. Acta {\bf 56} (1983) 75}

\REF\jain{X.G. Wu, G. Dev, and J.K. Jain, Phys. Rev. Lett. {\bf 71}
(1993) 153}

\REF\clark{R.G. Clark, {\it et al.}, Phys. Rev. Lett. {\bf 60}
(1988) 1747; {\bf 62} (1989) 1536}

\REF\maksym {P.A. Maksym, J. Phys.: Cond. Mat. {\bf 1} (1989) 6209}

\REF\mcdonald {I.A. McDonald, Ph.D. thesis, Princeton University;
I.A. McDonald and F.D.M. Haldane, in preparation}

\REF\symm{C. Nayak and F. Wilczek, PUPT 1515, IASSNS 94/109}

\REF\jain{J. Jain, Phys. Rev. Lett. {\bf 63} (1989) 199;
Phys. Rev. {\bf B40} (1989) 8079; {\bf B41} (1990) 7653.}

\REF\fwg{F. Wilczek, in {\it Fractional Statistics and Anyon
Superconductivty}, {\it ibid}. pp. 80-88; M. Greiter and F. Wilczek,
Mod. Phys. Lett. {\bf B4} (1990) 1063.}

\REF\fixedpoint{C. Nayak and F. Wilczek, Nucl. Phys. {\bf B 430}
(1994) 534}

\REF\goldman{V. Goldman, B. Su, and J. Jain, Phys. Rev. Lett. {\bf 72}
(1994) 2065}

\REF\drifts{R. Levien, C. Nayak, and F. Wilczek,
PUPT 1516, IASSNS 94/108}

\REF\hlr{B. Halperin, P. Lee, and N. Read, Phys. Rev. B {\bf 47}
(1993) 7312}

\REF\bonesteel{N.E. Bonesteel, Phys. Rev. {\bf B 48} (1993) 11484}

\REF\moon{K. Moon, {\it et al.}, Phys. Rev. Lett. {\bf } (1994) }

\chapter{Introduction}

There is now convincing experimental and numerical evidence
that a spin-singlet fractional quantum Hall state at $\nu=2/3$ is
realized at low densities in extremely pure samples.
The spin-singlet state is of fundamental interest because
its spin-symmetry is an emergent property (the microscopic theory
is not accurately symmetric) that arises
from special features of the correlated wavefunction.
This symmetry is not even approximately valid in the
generic case (eg. at other filling fractions) and is
not manifest in the standard effective theory.
Such a possibility was first raised
by Halperin [\halperin], who noted long ago that band mass and $g$-factor
corrections make the ratio of Zeeman to cyclotron energies
$\sim {1\over 60}$. Correlations can compete with this
small spin-dependent energy. The $\nu=2/3$ state is the simplest
fractional quantum Hall state which takes advantage of this circumstance.
A simple trial wavefunction for this state has been proposed
by Wu, Dev, and Jain [\jain]. This wavefunction has a large
overlap with the ground state found by numerical diagonalization
of the Hamiltonian for small numbers of particles[\jain,\mcdonald,\maksym].
In the tilted-field experiments of Clark, {\it et al.} [\clark],
which were crucial to the identification of the spin-singlet character
of this state, an in-plane magnetic field was introduced,
thereby increasing the Zeeman energy but leaving the cyclotron
energy unchanged. As the Zeeman energy was increased to a critical
value, the quantum Hall state was destroyed. At still higher
Zeeman energies, a quantum Hall state reappeared.
This is nothing but the transition from spin-singlet to
spin-polarized states as a function of the Zeeman energy.
In this paper, we will study that transition in detail.

In an earlier paper [\symm], we noted that the $K$-matrices
which describe the spin-polarized and spin-singlet states
are precisely the same. As a result, we argued, the boundaries
between regions of different phases in a first-order phase
transition, or a second-order phase transition in the presence
of disorder, will not support gapless excitations.
Since these states are
so similar, one might actually wonder whether there has to be
a phase transition at all. However, despite
the identity of their $K$-matrices,
the states are distinguished by
other topological quantum numbers.
For instance,
the polarized state has shift $S={1\over\nu}N-{N_\phi}=0$
on the sphere while the singlet state has shift $S=1$.
Still, one might
doubt that the gap goes quite to zero since, as we noted above,
there are no
gapless excitations at phase boundaries. Experiments are unclear
on this point since the non-vanishing $\rho_{xx}$
which they find could be indicative of a small
but non-vanishing gap. In any case, the possibility of metallic states
at the transition needs elucidation, whether or not they survive to $T=0$.
To address these questions, we exploitflux-trading arguments to
relate the $\nu=2/3$ transition to the $\nu=2$ transition
(as we suggested in our earlier paper [\symm]). At $\nu=2$, in the
vicinity of the transition, there are two nearly degenerate
Landau levels. We then have, effectively, a two-component
Hall system, similar to those analyzed in the context of
double-layer or single-wide layer systems. We relate the
transition at $\nu=2$ to a ferromagnetic transition
as a function of magnetic field below the critical temperature
of such a model. In this picture, the charge transfer gap
remains non-vanishing. Finally, we use flux-trading arguments
again to construct a
non-Fermi liquid effective field theory for a possible
finite-temperature metallic
state at the transition and describe its phenomenology.
Some of the effects we suggest appear to have been observed
in existing experimental and numerical work [\clark,\mcdonald].
If further work
confirms this, and especially if the predicted
metallic state could be demonstrated, it would
form an impressive demonstration of the
fruitfulness of the flux-trading concept.

\chapter{Relating $\nu=2/3$ to $\nu=2$}

The states of interest at $\nu=2/3$ are both related to
integer quantum Hall states at $\nu=2$ by
adiabatically trading magnetic flux for statistical flux.
In this familiar procedure, we simultaneously
change the statistics of the particles and the magnetic
field according to $n\Delta\Bigl({\theta\over\pi}\Bigr)=\Delta B$,
thereby leaving the system unchanged in the mean-field approximation
[\jain,\fwg].
The change of statistics may be implemented with a Chern-Simons field
which carries an average flux $\Delta B$. If we continue
this procedure until $\Delta\theta=2\pi p$, where $p$ is
an integer of either sign, then the statistics
is again fermionic. This is a state
of electrons in field ${B_{eff}}=B+\Delta B$, or, in other words, at
inverse filling fraction ${1\over{\nu'}}={1\over\nu}+2p$, which,
in mean-field approximation, is identical to the state at filling
fraction $\nu$. The mean-field approximation is not exact,
of course, so there are still gauge field
fluctuations to be dealt with -- ie. the Chern-Simons field
is not really equal to its mean -- but these shouldn't be
{\it qualitatively} important because both states have a gap. In
particular, starting with a state at $\nu=2$ and decreasing the
$B$ field by two flux tubes per electron, we arrive at the
state at $\nu=2/3$, which should be qualitatively similar.
Later, we will discuss the quantitative differences resulting from
the gauge field fluctuations.

An ambiguity arises because there are,
in fact, two possible integer quantum Hall states at
$\nu=2$. If the cyclotron energy is less that the Zeeman
energy, ${E_c}<{E_Z}$, then the first and second Landau
levels with spin aligned along ${\bf B}$ are filled.
If, however, ${E_c}>{E_Z}$, the first spin-aligned Landau
level and the first spin-reversed Landau level are filled;
this is a spin-singlet state. As we mentioned in the introduction,
the latter case is realized in GaAs-AlAs systems and the
observed state at $\nu=2$ is a spin-singlet. When an in-plane
magnetic field, ${B_\parallel}={B_\perp}\tan\theta=B\sin\theta$,
is turned on, ${E_Z}\propto B$ increases while ${E_c}\propto {B_\perp}$
is held fixed. At $\nu=2$, band mass and $g$-factor corrections
are so large that an enormous in-plane magnetic field would be necessary
to favor the spin-polarized state, ie. $\cos\theta\sim{1\over60}$.
At $\nu=2/3$, however,
the aforementioned gauge field fluctuations can have the
important quantitative effect of renormalizing the mass,
thereby decreasing the cyclotron energy considerably.
To get an idea of the magnitude of the mass renormalization,
we can simply look at the measured quasiparticle gap
at $\nu=2/3$. First, it is important to remember that
the quasiparticle gap is not the cyclotron energy,
${E_c}={{eB_{eff}}\over{m_r}}$, but the difference between
cyclotron and Zeeman energies,
$${E_g}={{eB_{eff}}\over{m_r}}-{{geB}\over{m_e}}\eqn\gap$$
because the lowest energy excitations are precisely
from the lowest spin-reversed Landau level to the
second spin-aligned Landau level. (Note that the
second term has $B$ rather than $B_{eff}$ because the Zeeman
coupling is unaffected by the flux-trading procedure.)
Fitting this expression to the measured gap at $\nu=2/3$,
we can obtain $m_r$. This value of the renormalized mass
may be used to estimate the tilted-field angle at which
the Zeeman and cyclotron energies are the same and the transition
occurs:
$${{e{B_{eff}}\cos\theta}\over{m_r}}={geB\over{m_e}}\eqn\transangle$$
or, simply, $\cos\theta = 3g({m_r}/{m_e})$. Experiments indicate
${m_r}\approx 10{m_b}$, so we find $\cos\theta\sim 0.5$.

In other words, we have estimated the cyclotron energy to
be of the same order of magnitude as the Zeeman energy,
in agreement with the experiments of
Clark, {\it et al.}, who find a transition at $\cos\theta\sim 0.9$.
This estimate was extremely crude, but it
is possible that a careful
calculation of gauge-field corrections to the
effective mass, perhaps taking into account disorder,
could predict this value more accurately.
However, if such a calculation fails to accurately
locate the transition, then the discrepancy might be due to a more fundamental
problem. The flux-trading arguments depend for their validity
on the existence of a gap, or at least a Fermi surface.
We assume that this condition is met.
While it is a logical possibility that the gap does go to zero
and a Fermi surface does not develop, in which case
gauge-field fluctuations can completely destabilize the
mean-field theory and we cannot relate the physics at
$\nu=2$ to the physics at $\nu=2/3$, we believe that the
semi-quantitative consistency of our description
makes such a possibility unlikely.

Given the assumptions and qualifications just expressed,
we are led to anticipate that the
basic physics of the $\nu=2/3$ states be visible in the
$\nu=2$ system, which is {\it a priori} easier to understand.
We will adopt this point of view here.

\chapter{Physical Picture at $\nu=2$}

Let's now consider the $\nu=2$ system in the vicinity of the
transition. The second spin-aligned and the first spin-reversed
Landau levels are nearly degenerate (the filled first spin-aligned
Landau level is essentially inert; we ignore these electrons
here and in what follows). Naively, there will be many low-energy
excitations because there are twice as many available states
as there are electrons. Hence, interactions will be crucial
even at integer filling fraction. This system is very
similar to a double-layer or single wide quantum well
system at $\nu=1$ (again, ignoring the inert
filled Landau level). The two Landau levels play the role of
the two layers. Tuning the magnetic field away from the transition
is analogous to unbalancing the two wells.
The strong Coulomb repulsion leads to
a ferromagnetic exchange interaction
($J\sim{{e^2}\over{\epsilon{l_H}}}{\buildrel{>}\over\sim}10K$)
which will order the
spins at low temperature. There are then two
possibilities, depending on whether inter-level or intra-level
interactions are stronger (the latter are typically
stronger in real double layer systems, of course).
\vskip 2cm

{\it Ising case.} When the inter-level interactions are
stronger than intra-level interactions, it is energetically favorable
to have all of the electrons in the lower of the two Landau
levels (if inter-layer interactions were weaker, then
it might be energetically favorable to have some
electrons in the higher level in order to decrease the interaction
energy). At the transition, the electrons all switch into
the other level. Let us consider, for a moment, the Ising model at
zero temperature. In the Ising model, the spins are all aligned
along the field. As the field is reversed, the spins become
reversed. This is a first-order transition (with a
possibility of hysteresis). At finite temperature, there
is a transition to a paramagnetic phase. Returning, now, to the
case of two nearly degenerate Landau levels we see the
following analogy. The two low-temperature ferromagnetic
phases of the Ising model (spin up and spin down) are
analogous to the two states of electrons at $\nu=2$. These
are related by a $Z_2$ symmetry which is broken in the
Ising model by a non-zero field and in the $\nu=2$ system
by a non-zero splitting between the two levels in question.
If we hypothesize that both of these $Z_2$ models are in the
same universality class, then we would predict a second-order phase
transition at finite temperature which terminates
the line of first-order phase transitions.
To observe it, one could tune to the first-order phase
transition at low-temperature and then raise the temperature.
Of course, the Coulomb interactions set the
scale for this transition temperature, so the $\nu=2/3$ state
would probably be destroyed before this temperature is reached.
If there were a separation of scales, however, and this
transition could be observed, then we would observe
a logarithmically diverging specific heat, a power law
spin-polarization, and the other characteristic behaviors of
the Ising universality class.

At a first-order phase transition point, there can be phase
coexistence. At the boundaries between the two phases
(regions of spin-up or spin-down in the Ising
analogy), one
might expect gapless edge modes. As we pointed
out in [\symm], this expectation is seen to
be incorrect, once inter-edge interactions are taken
into account. Let now us analyze this more carefully.
At $\nu=2$, the $K$-matrix describing
excitations at the boundary between two
phase regions is simply:
$$K = \left(\matrix{1&0&0&0\cr 0&1&0&0\cr
0&0&-1&0\cr 0&0&0&-1}\right)\eqn\kbound$$
At $\nu=2/3$, it is
$$K =  \left(\matrix{1&2&0&0\cr 2&1&0&0\cr
0&0&-1&-2\cr 0&0&-2&-1}\right)\eqn\kboundtt$$
In both cases the electron tunneling
operator can be a relevant operator when
inter-edge interactions are sufficiently strong. (In the former
case, the tunneling operator is relevant even in the
absence of these interaction.)
Consider the effective Lagrangian for these edge modes:
$${S_0} =  \int dx\,d\tau\,
\Bigl({K_{ij}} {\partial_\tau} {\phi^i} {\partial_x} {\phi^j}
  ~+~ {V_{ij}} {\partial_x} {\phi^j} {\partial_x} {\phi^j}\Bigr)
\eqn\llaction$$
At $\nu=2$, $K_{ij}$ is given by \kbound\ and the electron tunneling
operators of interest are $e^{i({\phi^1}-{\phi^3})}$
and $e^{i({\phi^2}-{\phi^4})}$. The first tunneling operator
corresponds to tunneling between the lowest spin-aligned
Landau level on either side of the phase boundary. The
second tunneling operator corresponds to tunneling between the lowest
spin-reversed Landau level and the second spin-aligned Landau level.
If a gap forms, it is presumably because these pairs of modes
form gaps. The tunneling operator
$${S_t} = \int d\tau\,{dx}\, \Bigl(\,{te^{i{k_{13}}x}}
\,{e^{i({\phi^1}-{\phi^3})}} + c.c.\Bigr)\eqn\etunlag$$
(and  the same operator with  $1\rightarrow2$ and $3\rightarrow4$)
can lead to the formation of a gap if it exists.  However,
this is a momentum non-conserving process unless ${k_{13}}=0$.
This will not be a problem if there are umklapp processes
at wavevector $k_{13}$, but in general there won't be, so
we will need ${k_{13}}=0$. This is just the statement that the two
edges be separated by a distance less than a magnetic
length because $k_{13}$ is equal to the magnetic flux
passing through the region between the two edges.
This condition is almost certainly satisfied by the edges
of the two lowest Landau levels on either side of the boundary
(the lowest Landau level is the same in both phases, of course). It
is highly plausible that this is also true of the edges
of the second Landau levels (which are different in the two phases),
at least once the lowest levels have paired off and formed
a gap. However, it is possible that there is some aspect
of the microscopic physics which distinguishes between
the lowest and second Landau levels. Such effects could prevent
${k_{13}}=0$ from being satisfied, but we will assume
that this does not happen. Numerical investigations might
shed valuable light on this issue.

The second condition that must be met in order for
a gap to form is that the tunneling operator \etunlag\
be relevant (ie. have positive dimension in momentum space).
The scaling dimension of the tunneling
operator may be obtained from the scalar field two-point
function. To obtain the scalar field two-point function, we
diagonalize the Lagrangian \llaction.  The result of these
calculations is that the scaling dimension of the operator
$${S_t} = \int d\tau\,{dx}\, \Bigl(\,{t}
\,{e^{i{l_i}{\phi^i}}} + c.c.\Bigr)\eqn\egentunlag$$
is $2-{\Delta_l}$,
where ${\Delta_l}= {1\over 2}\,{a_\mu^i}{a_\mu^j}{l_i}{l_j}$.
Here, the $a_\mu$'s are the simultaneous eigenvectors
of the $K_{ij}$ and $V_{ij}$ matrices of eq. \llaction;
they are orthonormal with respect to $K_{ij}$. ${\Delta_l}$
satisfies the inequality:
$${\Delta_l}\geq{1\over2}{K^{-1}_{ij}}{l_i}{l_j}\eqn\dkimpineq$$
Hence, the maximum scaling dimension of the tunneling operator
\egentunlag\ is $2-{1\over2}{K^{-1}_{ij}}{l_i}{l_j}$.

Actually, inter-edge interactions are unnecessary at $\nu=2$;
the tunneling operator \etunlag\ is a dimension 1 operator
even when $V_{ij}$ is diagonal. Non-zero off-diagonal elements
of $V_{ij}$ can make the tunneling operator even more relevant,
up to a maximum dimension of 2, according to the arguments of
the previous paragraph. At $\nu=2/3$, however, these interactions
are necessary; in their absence, \etunlag\ is an irrelevant
operator. However, ${K^{-1}_{ij}}{l_i}{l_j}=0$, so sufficiently
strong inter-edge interactions can make this operator relevant,
again to a maximum dimension 2. Hence, we find that so long as
the edges of the different phase regions are in close proximity
-- close enough to make ${k_{13}}=0$ and to make inter-edge
interactions strong enough to make \etunlag\ relevant --
these phase boundaries do not support gapless excitations.
\vskip 2 cm

{\it XY case.} When inter-level interactions are weaker than intra-level
interactions, it is energetically favorable, near the
transition, to excite some electrons to the higher level
because the interaction energy is thereby lowered. There
is a critical level-splitting at which this begins to
occur. When this happens, the spin vector rotates away
from the $z$-axis into the $x-y$ plane (the spin remains
maximal as a result of the strong ferromagnetic exchange
interaction). The entire system, including the electrons
in the lowest Landau level, is then neither a
spin-singlet nor a spin-polarized state.
Hence, the transition occurs in two steps in this
scenario: spin-singlet $\rightarrow$ partially polarized
$\rightarrow$ fully polarized.

This may be described phenomenologically with the following
effective Hamiltonian.
$$H = {S_x^2} + {S_y^2} + \eta{S_z^2} - h{S_z} + \ldots\eqn\effham$$
The ellipses indicate gradient and higher-order terms.
Here, $\eta>1$ is a measure of the anisotropy between
inter- and intra-layer interactions. For
$\eta<1$, \effham\ describes the Ising case
we discussed earlier.  $h$ is a measure
of the splitting between the two levels. We have neglected
the term $({S^2}-{S^2_{\rm max}})^2$ (where ${S_{\rm max}}$ is $N/2$,
and N is the number of electrons excluding the ``inert''
electrons in the lowest Landau level)
which leads to symmetry breaking,
but at low temperature this term just enforces the
constraint ${S^2}={S^2_{\rm max}}$. Substituting this constraint
into \effham, we find:
$$H = (\eta-1){S_z^2} - h{S_z} + {\rm const.} + \ldots\eqn\modham$$
The minimum of this Hamiltonian is at
${S_z}={\rm min}({h\over{2(\eta-1)}}, {S_{\rm max}})$. In other words,
when the splitting between levels is large compared to
the anisotropy in interaction strengths the system
is either in a fully spin-polarized state or a spin-singlet.
When the splitting is small, the system is partially polarized.
Clearly, the order parameter, $S_z$, is continuous, although its
first derivative is not. Hence, the spin-singlet to partially
polarized and partially polarized to fully polarized transitions
are both second-order phase transitions transitions.

As has been discussed in the context of double-layer
systems, the partially polarized state has a $U(1)$ symmetry
corresponding to rotations in the ${S_x}-{S_y}$ plane.
The $U(1)$ charge is just ${n_\uparrow}-{n_\downarrow}$,
the number of spin-up electrons minus
the number of spin-down electrons. This symmetry is broken
at $T=0$, and there are Goldstone bosons. At finite temperature,
there is a Kosterlitz-Thouless phase transition.
The most natural wavefunction for the partially polarized
state is the $(1,1,1)$ wavefunction for a two-component
Hall system. Some modification of this wavefunction is required,
however, because one of the ``layers'' is actually a second
Landau level.

As in the Ising case, the phase boundaries between
spin-singlet or fully polarized states and the partially
polarized state, which can occur in the presence of disorder,
do not support gapless excitations. The arguments are completely
analogous to those we used in that case. This is, in fact,
quite intuitive. The transitions we are considering are
changes of the spin configuration of the state. The charge transfer
gap may become small, but it does not close. One might worry that
the closing of the gap, due to the existence of Goldstone bosons
in the partially polarized state, invalidates the flux-trading procedure.
However, the charged modes -- which are the only modes affected
by the flux-trading procedure -- are not gapless, so we
expect that this does not occur.
\vskip 2 cm

We have presented two scenarios for the spin-singlet to
spin-polarized phase transition at $\nu=2/3$. It is natural
to ask whether one or the other of these possibilities
is better suited to describing experiments. Numerical
studies indicate that the the phase transition at
$\nu=2/3$ is a first-order phase transition with a
simple level crossing [\mcdonald]. Experiments indicate that the
transition is directly from a spin-singlet to a
spin-polarized state; there is no evidence of a
partially polarized state [\clark]. Thus, if our flux-trading
arguments are correct, the Ising-type transition
analyzed at the beginning of this section describes
the $\nu=2/3$ transition. However, experiments at
$\nu=4/3$ -- the particle-hole conjugate of $\nu=2/3$ --
find a two-step transition with a partially
polarized state at the intermediate step [\clark]. This
transition appears to be described by the $XY$ picture
analyzed in the latter part of this section.
However, experiments find a non-zero $\rho_{xx}$, indicative of a small or
vanishing gap, at the $\nu=2/3$ and $4/3$ transitions.
Presumably, the experiments are conducted
at temperatures higher than a possible gap, so
the state at the transition appears to be metallic.

\chapter{A Possible Non-Fermi Liquid Metallic State}

Using flux-trading arguments yet again, we can construct
a mean-field theory with a Fermi surface for the state
at the transition. Gauge-field fluctuations
lead to corrections of a non-Fermi liquid form, but they do not
destabilize the Fermi surface. This is completely analogous to the
procedure used by Halperin, Lee, and Read [\hlr]
to construct the non-Fermi liquid
theory of the half-filled Landau level. The new feature
is the degeneracy between two Landau levels. As a result, we have
a non-Fermi liquid theory with spin.

Of course, this mean-field theory is only an ansatz. If it
is not energetically favored compared to the possible
incompressible states, which is what we expect, then it will
not be realized at $T=0$. In particular, we will find that there
is a marginally relevant operator in the effective field
theory of this model which is a special feature of models
with spin. This operator leads to the formation
of a gap and condensation at low temperature. The low-temperature
state is presumably the state which we discussed in the
previous section. At temperatures above the transition
temperature, we expect a non-Fermi liquid metallic state, however.

The starting point is two degenerate Landau levels -- the second
spin-aligned level and the first spin-reversed level -- and
one electron per magnetic flux tube. In this mean field theory,
we will assume that half of the electrons are in each
Landau level. We then introduce two Chern-Simons gauge
fields which are coupled to the electrons in each of the two
levels. These gauge fields attach two flux tubes to each electron
in the direction antiparallel to the magnetic field. As usual,
these gauge fields have no effect; in particular, they
do not change the statistics of the
electrons. However, in mean field theory,
the system is now one of electrons with spin in zero net field.
Of course, the gauge-field fluctuations cannot be neglected
because the mean-field theory is gapless. Elsewhere [\fixedpoint], we
have used the renormalization group to tame the gauge-field
fluctuations in the context of the half-filled Landau
level. The same framework may be applied to this problem, but with
a few twists.

We begin with the effective action:
$$\eqalign{S &=
\int\,{d\omega\,{d^2}k \biggl\{{\psi_\sigma^{\dagger}}\bigl(i\omega
- \epsilon(k)\bigr){\psi_\sigma}\biggr\}}
+ \int\,{d\omega\,{d^2}k\, {a_{\sigma0}}\epsilon_{ij}{k_i}
{a_{\sigma j}}}\cr &\qquad
+ g \int\, {{d\omega}\,{d\omega'}\,{d^2}k\,{d^2}q
\biggl\{{\psi_\sigma^{\dagger}}(k+q,\omega+\omega'){\psi_\sigma}(k,\omega)
\Bigl({a_{\sigma i}}(q,\omega'){\partial\over{\partial{k_i}}}\epsilon(q+2k) +
{a_{\sigma 0}}(q,\omega')\Bigr)\biggr\}}\cr &\qquad
+ {V_0}\int\,{d\omega}\,{d\omega'}\,{d\omega''}\,
{d^2}k\,{d^2}k'\,{d^2}k'' {\psi_\sigma^{\dagger}}(k+k',\omega+\omega')
{\psi_\sigma}(k',\omega')\, {1\over{k^x}}\,\times\cr &\qquad
{\psi_{\sigma'}^{\dagger}}(-k+k'',-\omega+\omega'')
{\psi_{\sigma'}}(k'',\omega'')\cr}\eqn\nfleffac$$
${\psi_\sigma}$ are the fermion fields; $\sigma$ is the spin or,
alternatively, the Landau level index. The $a_\sigma$ are
the gauge fields coupled to the two fermion fields. The final term
is a four-fermion interaction representing electron-electron
interactions. If $x=1$, it is just
the Coulomb interaction, but we will let $x$ be arbitrary
for now. It will
be convenient to rewrite the action in terms of the gauge
fields ${a_c}={1\over2}({a_\uparrow}+{a_\downarrow})$ and
${a_s}={1\over2}({a_\uparrow}-{a_\downarrow})$. Furthermore, we
will need to use the $a_{\sigma 0}$ equation of motion,
or Chern-Simons constraint:
$$\epsilon_{ij}{k_i}{a_{\sigma j}(k) = g \int\,{d^2}q\,{d\omega'}\,
{\psi_\sigma}^{\dagger}}(k+q,\omega+\omega')
{\psi_\sigma}(q,\omega)~.\eqn\constr$$
(no sum over $\sigma$). If we substitute this constraint back
into the non-local
four-fermion interaction, the important role
of this interaction becomes clear.  It takes the form
$$S_a~=~\int\,{d\omega}\,{d^2}k\,\epsilon_{ij}\epsilon_{mn}
{k_i}{k_m}k^{-x}{a_{cj}}(k,\omega){a_{cn}}(-k,-\omega)~.\eqn\keytermac$$
Thus as $x$ is increased, the long-range fluctuations of the gauge
field $a_c$ are suppressed. The four-fermion interaction does not affect
the gauge field $a_s$. However, there should also be terms in
\nfleffac\ representing spin-spin interactions which will lead
to a term of the form of \keytermac\ but with $x=0$ since
spin-spin interactions are local. Even otherwise, such a term
will arise at one-loop anyway, so we should include such a term in the
action. Introducing renormalization counterterms and using
a regularization procedure analogous to dimensional regularization,
as in [\fixedpoint], we have the action:
$$\eqalign{S &=
\int\,{d\omega\,{d^2}k {\psi_\sigma^{\dagger}}\bigl(iZ\omega
- Z{Z_{v_F}}\epsilon(k)\bigr){\psi_\sigma}}
\,+\, \int\,{d\omega\,{d^2}k {a_{\sigma 0}}
{\epsilon_{ij}}{k_i}{a_{\sigma j}}}\cr &\qquad
+ \int\,{d\omega}\,{d^2}k\,\epsilon_{ij}\epsilon_{mn}
{k_i}{k_m}k^{-x}{a_{cj}}(k,\omega){a_{cn}}(-k,-\omega)\cr &\qquad
+ \int\,{d\omega}\,{d^2}k\,\epsilon_{ij}\epsilon_{mn}
{k_i}{k_m}{a_{sj}}(k,\omega){a_{sn}}(-k,-\omega)\cr &\qquad
+ {{\mu}^{{1-x}\over{2}}}
{g_c}{Z_{g_c}} \int\, {{d\omega}\,{d\omega'}\,{d^2}k\,{d^2}q\,
{\psi_\sigma^{\dagger}}(k+q,\omega+\omega'){\psi_\sigma}(k,\omega)
\Bigl({a_{ci}}(q,\omega'){\partial\over{\partial{k_i}}}\epsilon(q+2k) +
{a_{c0}}(q,\omega')\Bigr)}\cr &\qquad
+{{\mu}^{{1-x}\over{2}}}
{g_s}{Z_{g_s}} \int\, {{d\omega}\,{d\omega'}\,{d^2}k\,{d^2}q
\,{\psi_\sigma^{\dagger}}(k+q,\omega+\omega'){\tau^3_{\sigma\sigma'}}
{\psi_{\sigma'}}(k,\omega)
\Bigl({a_{si}}(q,\omega'){\partial\over{\partial{k_i}}}\epsilon(q+2k) +
{a_{s0}}(q,\omega')\Bigr)}\cr}\eqn\impeffac$$
$a_c$ couples to the total charge while $a_s$ couples to the $z$-component
of the quasiparticle spin; ${\tau^3_{\sigma\sigma'}}$ is the
appropriate Pauli matrix. The action \impeffac\ is not
$SU(2)$ invariant because the magnetic field determines a
preferred quantization axis. There is no quadratic coupling
of $a_s$ to $a_c$; such a term will not be generated in perturbation
theory because the contributions of spin-up and spin-down
quasiparticles will cancel.

In [\fixedpoint], we showed that the fermion-gauge field interaction
is relevant for $x<1$ and found a non-Fermi liquid fixed point
for $1-x$ small. At $x=1$, we found logarithmic corrections
to Fermi liquid behavior. We hypothesized that this fixed point exists
even at $x=0$. A number of authors have claimed to have constructed
this fixed point using resummations of perturbation theory.
Here we have an example of fermions interacting with two
gauge fields, one of which has $x=1$, the other $x=0$.
We expect that for most correlation functions the
gauge field coupling to $S_z$ with $x=0$ will dominate
because it has the more singular low-energy behavior,
so we ignore the other gauge field in what follows.

Adopting the results of [\fixedpoint], we have the
$\beta$-functions for the coupling constant,
${\alpha_s}={{{g_s^2}{v_F}}\over{(2\pi)^2}}$:

$$\beta({\alpha_s}) = -{1\over 2}(1-{x_s})\,{\alpha_s} + 2{\alpha_s}^2 +
 O({\alpha_s^3})\eqn\bfcns$$
and the anomalous dimensions,
$$\eta_{v_F}({\alpha_s}) = -\eta({\alpha_s}) =
-2{\alpha_s} + O({\alpha^2})~.\eqn\anomdim$$
This leads to the anomalous scaling form for the fermion
two-point function:
$${G^{(2)}}(\omega,{v_F}{r},\alpha,\mu) =
{\omega}^{-1 + \eta}
{G^{(2)}}(1,{{{v_F}r}\over{\omega^{1+{\eta_{v_F}}}}},
{\alpha^*},\mu)\eqn\rgscalreltp$$

Such a non-Fermi liquid metallic state has dramatic experimental
consequences. Surface acoustic wave propagation will exhibit
anomalies at the Fermi wavevector $k_F$, which is given
by $\pi{k_F^2}={1\over6}B$ at $\nu=2/3$ (half the electrons
are inert and the remaining half are divided into spin-up
and spin-down). Furthermore, magnetic focusing experiments of the
type conducted by Goldman, {\it et al.} [\goldman] in the vicinity of
$\nu={1\over2}$ would reveal the existence of the Fermi
wavevector $k_F$. Time-of-flight measurements in such an experiment,
as discussed in [\drifts], will reflect the Fermi velocity
renormalization due to $a_s$.

As we mentioned at the beginning of this section, there
are new possibilities for pairing as a result of the
additional quantum number, spin, and its gauge field. This
is because spin-up and spin-down electrons are oppositely
charged with respect to the gauge field $a_s$. As Bonesteel
has observed in the case of double-layer systems [\bonesteel],
this gauge-field mediated interaction is attractive.
The enhanced pairing
interaction presumably leads to a finite temperature
transition to the incompressible state of the previous section.

We mention now briefly a fundamental and at first sight disturbing
point, that we will explore in reater depth elsewhere. We
have relied heavily, in our analysis, on coupling gauge
fields to $S_z$ -- a quantity that is not strictly conserved.
Local gauge invariance must be exact, however. What is going on here?

Local gauge invariance can be maintained, without its usual
implication of associated conservation laws, if (and
only if) {\it non-local} terms are allowed in the action.
Thus, for example in our context, spin-flip processes
must be accommodated by appropriate non-local interactions.

\chapter{Discussion}

In the preceding sections, we have applied the method of
adiabatic flux-trading and analogies with double-layer
systems to analyze the phase transition between spin-singlet
and spin-polarized states at the same filling fraction.
We argued that the phase transition at $\nu=2/3$
is qualitatively similar to that at $\nu=2$.
A simple way of testing the validity of
this picture would be to see how the critical
tilt angle varies with the $g$-factor (which can be altered
by changing the Al concentration in AlAs-GaAs systems).
Even at $\nu=2$, we found two possibilities.
In the first, the transition
between polarized and singlet states is first-order.
The transition at $\nu=2/3$ appears to be of this type.
At $\nu=4/3$, the other possibility is realized. Two
first-order phase transitions occur; first from a fully polarized
state to a partially polarized state and thence to
a spin-singlet. This rich phase structure is reminiscent of
that of the double-layer Hall states [\moon].

As in the double-layer systems [\bonesteel], there is
the possibility of a metallic state. This phenomenology of this state
is qualitatively similar to that of the half-filled Landau level.
Two important differences result, however, from the presence
of a second gauge field. First, the gauge field which couples to spin
leads to a pairing instability. Second, spin-flip processes
are non-local in terms of the low-energy quasiparticles.
This latter fact can result in a dramatically altered response
of the spin degrees of freedom.

\ack{C.N. would like to thank Ian McDonald for a discussion
of his results.}

\endpage

\refout

\end